\documentclass[preprint]{aastex}
\usepackage{epsfig,emulateapj5,apjfonts}
\newcommand       \be           {\begin{equation}}
\newcommand       \ee           {\end{equation}}
\newcommand       \Angstrom     {\,{\rm \AA}}          
\newcommand       \eV           {\,{\rm eV}\,}
\newcommand       \K            {\,{\rm K}}
\newcommand       \cm           {\,{\rm cm}}
\newcommand       \s            {\,{\rm s}}
\newcommand       \erg          {\,{\rm erg}}
\newcommand       \kms		{\,{\rm km \, s}^{-1}}

\newcommand       \nH           {n_{\rm H}}
\newcommand       \xH           {x_{\rm H}}

\newcommand       \urad         {u_{\rm rad}}
\newcommand       \uHab         {u_{\rm Hab}^{\rm uv}}

\newcommand       \gtsim        {\gtrsim}
\newcommand       \ltsim        {\lesssim}

\newcommand	\cpar		{\psi}	

\newlength{\fullfigurewidth}
\shorttitle{Recombination on Grains}
\shortauthors{Weingartner \& Draine}

\begin{document}

\title{Electron-Ion Recombination on Grains and Polycyclic Aromatic 
Hydrocarbons}

\author{Joseph C. Weingartner}
\affil{CITA, 60 St. George Street, University of
Toronto, Toronto, ON M5S 3H8, Canada}
\email{weingart@cita.utoronto.ca}
\and

\author{B.T. Draine}
\affil{Princeton University Observatory, Peyton Hall,
        Princeton, NJ 08544, USA}
\email{draine@astro.princeton.edu}

\begin{abstract}

With the high-resolution spectroscopy now available in the optical and 
satellite UV, it is possible to determine the neutral/ionized column 
density ratios for several different elements in a single cloud.
Assuming ionization equilibrium for each element, one
can make several independent determinations of the electron density.  For 
the clouds for which such an analysis has been carried out, these different
estimates disagree by large factors, suggesting that some process (or
processes) besides photoionization and radiative recombination might play
an important role in the ionization balance.  One candidate process is 
collisions of ions with dust grains.

Making use of recent work quantifying the abundances of polycyclic aromatic
hydrocarbon molecules and other grains in the interstellar medium, as well
as recent models for grain charging, we estimate the grain-assisted ion 
recombination
rates for several astrophysically important elements.  We find that
these rates are comparable to the rates for radiative recombination for 
conditions typical of the cold neutral medium.  Including grain-assisted ion 
recombination in the ionization equilibrium analysis leads to increased
consistency in the various electron density estimates for the gas along
the line of sight to 23 Orionis.  However, not all of the
discrepancies can be eliminated in this way; we speculate on some other 
processes that might play a role.
We also note that grain-assisted recombination
of H$^+$ and He$^+$ leads to significantly lower electron fractions than 
usually assumed for the cold neutral medium.  

\end{abstract}


\section{Introduction}

The electron number density $n_e$ in diffuse interstellar clouds is often 
inferred from the observed 
column densities of different ionization
stages of metals.  The assumption of ionization equilibrium yields
\be
\label{eq:ionization_equil}
\Gamma({\rm X}^i) n({\rm X}^i) = \alpha_r ({\rm X}^{i+1}, T) n_e 
n({\rm X}^{i+1})~~~,
\ee
where $n({\rm X}^i)$ is the number density of the $i$-th 
ion of element X, $\Gamma({\rm X}^i)$ is the ionization 
rate for ${\rm X}^i\rightarrow{\rm X}^{i+1}$,
and $\alpha_r ({\rm X}^{i+1}, T)$ is the rate
coefficient (which depends on the gas temperature $T$) for 
recombinations ${\rm X}^{i+1}\rightarrow{\rm X}^i$.
It is usually assumed that radiative recombination
is dominant at low temperatures; dielectronic recombination is often 
important when $T \gtsim 10^4 \K$.  

The Goddard High Resolution Spectrograph on the {\it Hubble Space Telescope}
has made possible the simultaneous determination of $n_e$ through analysis
of the ionization balance of several different metals.  In some cases, the
values of $n_e$ inferred from different species differ
by large factors (Fitzpatrick
\& Spitzer 1997; Welty et al.~1999), suggesting either that the ionization 
is far from a steady-state equilibrium or that another process (or
processes) besides photoionization and radiative recombination affects the 
ionization balance.  

One candidate process is ion recombination
via electron transfer from a grain or
molecule to the ion during a collision (Weisheit \& Upham 1978; 
Omont 1986; Draine \& Sutin 1987; Lepp et al.~1988).  
The ``removed'' ion might remain in the gas phase (in a 
lower ionization stage) or it might be depleted from the gas altogether, if it 
sticks to the grain.\footnote{Henceforth, we will include molecules in the 
	``grain'' population.}
The grain-assisted ion removal rate is expected to vary from species to
species since the possibility of charge transfer depends on the electron
affinity of the ion and the collision rate depends on its mass.

The collision rate also depends on the grain charge
and size; when the grain charge $Z \le 0$, the ratio of the collision cross
section to the geometric cross section is larger for smaller grains 
(Draine \& Sutin 1987).  Thus, a large population of molecules or very small
grains with $Z \le 0$ could contribute significantly to ion removal.

Observations of emission from the diffuse interstellar medium
(ISM) have revealed 
features at 3.3, 6.2, 7.7, 8.6, and $11.3 \micron$ (see Sellgren 1994 for a 
review), which  have been identified as C\sbond H and C\sbond C 
stretching and bending 
modes in polycyclic aromatic hydrocarbons (PAHs; L\'{e}ger \& Puget 1984;
Allamandola, Tielens, \& Barker 1985).
Li \& Draine (2001) compared observations of diffuse Galactic emission with 
detailed model calculations for PAHs heated by Galactic starlight and found
that a C abundance\footnote{By ``abundance'' we mean the number of atoms of 
an element per interstellar H nucleus.} $\sim 4$--$6 \times 10^{-5}$ is 
required in hydrocarbon molecules with $\ltsim 10^3$ C 
atoms.\footnote{Electric dipole emission from this population can account for 
	the dust-correlated component of the diffuse Galactic microwave 
	emission (Draine \& Lazarian 1998; Draine \& Li 2001).}  
In recent charging models, PAHs are found to be predominantly neutral in
the diffuse ISM (Bakes \& Tielens 1994; Weingartner \& Draine 2001b).  

In \S \ref{sec:ion_destruct}, we estimate grain-assisted ion removal
rates for several metals, employing a grain size distribution that (a) 
includes PAHs in numbers sufficient to account for the observed
infrared and microwave emission 
and (b) reproduces the observed extinction of starlight (Weingartner \& 
Draine 2001a).  In \S \ref{sec:compare}, we compare the grain-assisted ion
removal rate with the radiative recombination rate, finding them to be
comparable in the cold neutral medium.  In \S \ref{sec:n_e}, we discuss the
determination of electron densities from observed column density ratios
when grain-assisted ion removal is important.  In \S \ref{sec:observations},
we apply these results to two observed lines of sight.  We find that the 
inclusion of grain-assisted ion removal cannot, on its own, reconcile 
the electron densities inferred from different species observed towards 
23 Ori.  In \S \ref{sec:additional_processes}, we speculate on additional 
processes that might affect ionization balance along this line of sight.
In \S \ref{sec:x_e}, we investigate the impact of grain-assisted ion 
recombination
on the electron fraction in the cold neutral medium.  We summarize
our conclusions in \S \ref{sec:conclusions}.

\section{\label{sec:ion_destruct} Ion Removal Rates}

Consider the collision of an ion X$^i$ with a dust grain of radius $a$
and charge $Ze$ ($e$ is the proton charge).  We assume that an electron is 
transferred from the grain to the ion if ${\rm IP}(a,Z) < 
{\rm IP}({\rm X}^{i-1})$, where ${\rm IP}(a,Z)$ is the grain ionization 
potential and ${\rm IP}({\rm X}^{i-1})$ is the ionization potential of 
X$^{i-1}$ (equal to the
electron affinity of X$^i$; see Table \ref{tab:ip_r_0} for our adopted   
values).  This is certainly an oversimplification since the electrostatic 
energy of the grain-ion combination usually increases following 
electron transfer.  However, 
it is not clear how to evaluate the 
electrostatic energy when the distance from the grain to the ion is $\ltsim$ 
a few angstroms.  Thus, our estimates for ion removal rates should be 
regarded as likely upper limits.  Below, we will estimate the extent to which 
these rates might be changed if the electrostatic energy changes were 
included in the analysis.  

We adopt the following expression for the ion removal rate:
\be
\frac{dn_i}{dt} = - \int dn_{\rm gr} \sum_Z f(a,Z) J_i(Z)
\, \theta[IP({\rm X}^{i-1})-IP(a,Z)]~~~,
\ee
where $n_i$ is the number density of X$^i$, $n_{\rm gr}(a)$ is the number
density of grains with radii $\le a$, 
$f(a,Z)$ is the fraction of the grains of radius $a$ that have charge $Ze$,
and $\theta(y) = 0$ if $y<0$ and 1 if $y \ge 0$.  Ions arrive at a grain
with size $a$ and charge $Z$ at the rate
\be
\label{eq:J_i}
J_i(Z) = n_i \left(\frac{8kT}{\pi m_p} \right)^{1/2} A_{\rm X}^{-1/2} \pi a^2
\tilde{J}(\tau_i \equiv akT/q_i^2, \xi_i \equiv Ze/q_i)~~~,
\ee
where $k$ is Boltzmann's constant, $m_p$ is the proton mass, $A_{\rm X}$ is 
the ion mass number, $q_i$ is the ion charge, and expressions for 
the dimensionless function
$\tilde{J}(\tau_i, \xi_i)$ are given in Draine \& Sutin (1987).  

Grain charge distributions in the diffuse ISM are set primarily by the balance
between electron loss via photoelectric emission versus electron gain via 
accretion from the gas.  
We calculate the steady-state charge distributions $f(a,Z)$ using
the charging algorithm given by 
Weingartner \& Draine (2001b) and the average interstellar radiation field 
(ISRF) spectrum for the solar neighborhood as estimated by Mezger, Mathis, \& 
Panagia (1982) and  Mathis, Mezger, \& Panagia (1983).  An often-used measure
of the radiation intensity is the parameter 
$G \equiv \urad^{\rm uv} / \uHab$, where $\urad^{\rm uv}$ is the 
energy density in the radiation field between $6 \eV$ and $13.6 \eV$
and $\uHab = 5.33 \times 10^{-14} \erg {\cm}^{-3}$ is the 
Habing (1968) estimate of the starlight energy density in this 
range.\footnote{
        For comparison, the interstellar radiation field estimated by
        Draine (1978) has $u=8.93\times10^{-14}\erg\cm^{-3}$ between
        6 and 13.6 eV, or $G=1.68$.
        }
For the ISRF, $G=1.13$.  Grain charging is largely determined
by the parameter 
\be
\cpar \equiv G\sqrt{T}/n_e
\ee
(Bakes \& Tielens 1994; Weingartner \& Draine 2001b), with only a weak
additional dependence on $T$.  
Examples of charge distribution functions $f(a,Z)$ are given in
Weingartner \& Draine (2001b).

\begin{figure*}[tbh]
\centerline{\epsfig{file=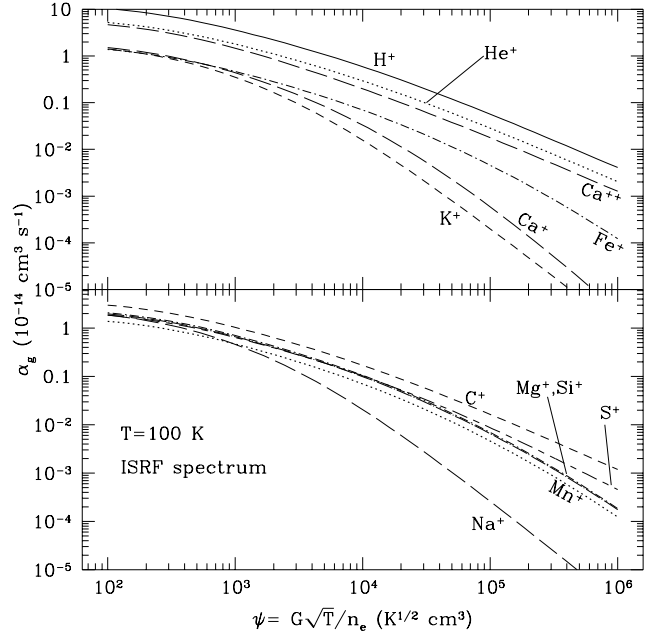,width=\fullfigurewidth}}
\caption{
\label{fig:alpha_g}
Rate coefficient $\alpha_g$ for grain-assisted removal of selected ions
as a function of the charging parameter $\cpar$, for the ISRF spectrum and gas
temperature $T=100\K$.
        }
\end{figure*}

We define a rate coefficient $\alpha_g({\rm X}^i, \cpar , T)$
for electron transfer to ion X$^i$ due to collisions with grains by
\be
\label{eq:alpha_g}
\frac{dn_i}{dt} = - \alpha_g({\rm X}^i, \cpar, T) n_i \nH~~~,
\ee
where $\nH$ is the H nucleus number density.
In Figure \ref{fig:alpha_g}, we display $\alpha_g$ as a function of $\cpar$ for
several ions X$^i$ and $T=100\K$.  In Figure \ref{fig:alpha_g}, and 
throughout this paper, we employ   
the grain size distribution from Weingartner \& Draine (2001a) with 
$R_V=3.1$ and $b_{\rm C}=6\times 10^{-5}$ (their favored distribution for
the average diffuse ISM).
Note that $\alpha_g$ decreases sharply as $\cpar$ 
increases, i.e., as the average
grain charge increases.  When $T\ltsim 10^3 \K$, the dependence 
of $\alpha_g$ on $T$ is quite mild since, for
neutral grains and $\tau_i \ll 1$, $\tilde{J} \propto T^{-1/2}$, cancelling
the factor of $T^{1/2}$ in equation (\ref{eq:J_i}).

For the cases illustrated in Figure \ref{fig:alpha_g}, we have also computed 
a modified ion removal rate coefficient $\alpha_g^{\prime}$, for which we
apply a different criterion for electron transfer:   
\begin{equation}
{\rm IP}(a,Z) - {\rm IP}({\rm X}^{i-1}) + \Delta U(Z,i) < 0 ~~~.
\end{equation}
Here, 
$\Delta U$ is the change in the interaction energy between the grain and
the ion due to the electron transfer.
We approximate (see Appendix)
\begin{equation}
\label{eq:DeltaU}
\Delta U(Z,z) \approx (z-Z-1)\frac{e^2}{(a+r_0)} + \frac{(2z-1) e^2 a^3}
{2 r_0 (a+r_0)^2 (2a+r_0)}~~~,
\end{equation}
where $Ze$ is the initial grain charge, $ze$ is the initial ion charge, and
$r_0$ is the atomic radius (see Table \ref{tab:ip_r_0} for our adopted values
of $r_0$).  Since the distance of closest approach between the ion and the 
grain surface is unknown, we have taken it to be equal to the atomic radius.  

The first term in equation (\ref{eq:DeltaU}) is the monopole contribution 
due to the net charges on the grain and ion.  Since the grains are 
predominantly neutral ($Z=0$) and most ions of interest are singly ionized
($z=1$), the inclusion of this term usually has a negligible effect on the
calculated ion removal rate.  The second term in equation (\ref{eq:DeltaU}) 
is due to the polarization of the grain by the ion.  Our classical 
calculation of the interaction energy assumes that the grain is (a) a 
perfect conductor and (b) a sphere with a perfectly sharp surface.  These
assumptions can lead to substantial errors in the estimate of the image 
potential when the ion-grain surface separation $\le$ a few angstroms.
We have neglected the dipole moment in the atom or ion induced
by the grain charge,
since the approximation of the atom/ion dipole as a point dipole
breaks down when the distance from the atom/ion to the grain surface is
comparable to the size of the atom---the case of interest here.

\begin{figure*}[tbh]
\centerline{\epsfig{file=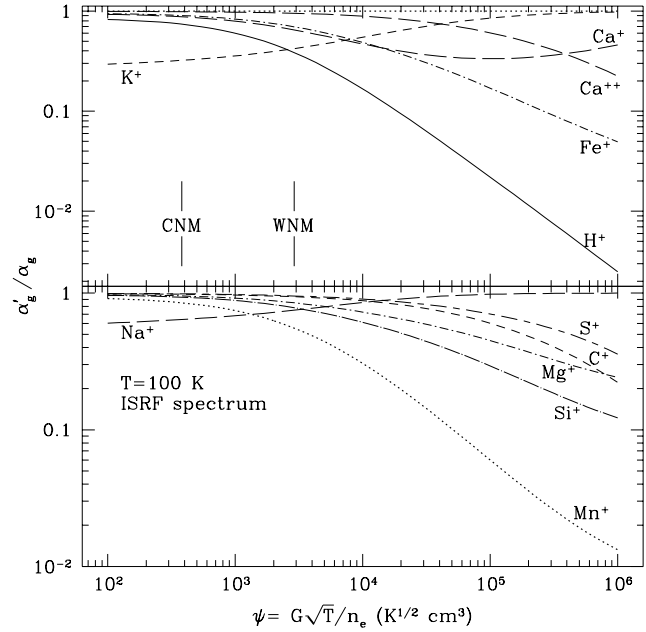,width=\fullfigurewidth}}
\caption{
\label{fig:compare}
$\alpha_g^{\prime}/\alpha_g$, where $\alpha_g^{\prime}$ is computed assuming
a more stringent criterion for the transfer of an electron from grain to ion,
in which the increase in electrostatic energy due to the interaction of the
grain and ion charge is taken into account in an approximate manner.  
Nominal cold and warm neutral medium conditions are indicated (``CNM'' and 
``WNM'', respectively). 
        }
\end{figure*}

In Figure \ref{fig:compare}, we display $\alpha_g^{\prime} / \alpha_g$.  
For cold and warm neutral medium conditions, $\alpha_g^{\prime} / \alpha_g$
is generally $\sim 1$; thus, it appears that, for most cases of interest, an 
accurate accounting of the increase in electrostatic energy following 
electron transfer would not substantially reduce the computed ion 
removal rate.

For a given ion, $\alpha_g$ can be represented approximately by the following
fitting formula:
\be
\label{eq:alpha_g_approx}
\alpha_g({\rm X}^i, \cpar, T) \approx \frac{10^{-14} C_0 \cm^3 \s^{-1}}
{1+C_1 \cpar^{C_2} \left[ 1+C_3 T^{C_4} \cpar^{-C_5 - C_6 \ln T} \right]}~~~,
\ee
with $T$ in kelvins and $\cpar$ in units of $\K^{1/2} \cm^3$; 
values of the fitting coefficients $C_0$---$C_6$ are given in 
Table \ref{tab:alpha_g} for selected ions.\footnote{A
	FORTRAN subroutine 
	implementing approximation (\ref{eq:alpha_g_approx}) is available on
	the World Wide Web at \url{www.cita.utoronto.ca/$\sim$weingart}.}  
Approximation 
(\ref{eq:alpha_g_approx}) is accurate to within $20\%$ 
when $10 \le T \le 10^3 \K$ and $10^2 \le \cpar \le 10^6 \K^{1/2} \cm^3$.
In the grain-assisted recombination of Ca$^{++}$, there is a possibility for 
two electrons to be transferred to the ion, resulting in Ca$^0$.
We assume that this occurs in a two-step process and that the energy
released following the first electron transfer is dissipated into heat, and
thus is not available to assist in the second transfer.  With these 
assumptions, the criterion for double electron transfer is
\be
{\rm IP}(a,Z+1) < {\rm IP}({\rm Ca}^0)  ~~~.
\ee
For this criterion, $10\leq T\leq10^3\K$, and 
$10^2\leq\psi\leq10^6 \K^{1/2} \cm^3$, the probability of double charge 
transfer can be approximated to within 10\% by
\be
f_{{\rm Ca}^{++} \rightarrow {\rm Ca}^0} \approx 
\frac{0.415}{1+ 6.38\times 10^{-5} \cpar^{1.252} (1+ 3.57 \times 10^{-6}
\cpar^{1.124})}~~~.
\label{eq:f_2}
\ee
Note, however, that the double electron transfer rate could be dramatically
reduced if electrostatic energy changes (as in eq.~[\ref{eq:DeltaU}])
are included in the criterion for electron transfer.  

\section{\label{sec:compare} Grain-assisted ion removal versus radiative 
recombination}

\begin{figure*}[tbh]
\centerline{\epsfig{file=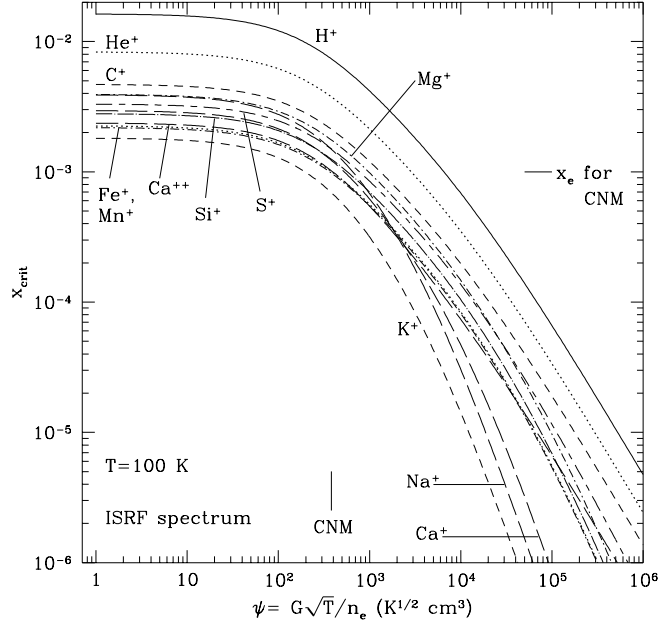,width=\fullfigurewidth}}
\caption{
\label{fig:x_crit_cnm}
Critical electron fraction:  if $x \le x_{\rm crit}$, then the indicated ion
is removed more effectively via collisions with grains than via 
radiative recombination.  The ticks labeled ``CNM'' are for canonical cold
neutral medium values of $\cpar$ and $x$.
        }
\end{figure*}

\begin{figure*}[tbh]
\centerline{\epsfig{file=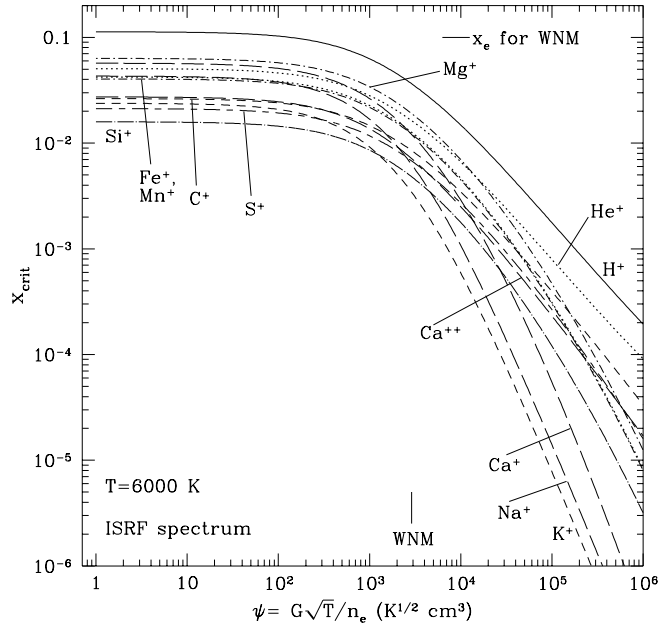,width=\fullfigurewidth}}
\caption{
\label{fig:x_crit_wnm}
Same as Figure \ref{fig:x_crit_cnm}, except for $T=6000 \K$ instead of 
$T=100 \K$.  The ticks labeled ``WNM'' are for canonical warm neutral 
medium values of $\cpar$ and $x$.
        }
\end{figure*}

To gauge the importance of grain-assisted ion removal in the diffuse ISM,
we plot in Figures \ref{fig:x_crit_cnm} and \ref{fig:x_crit_wnm} 
a critical electron fraction 
\be
x_{\rm crit}({\rm X}^i, \cpar, T) \equiv 
\frac{\alpha_g({\rm X}^i, \cpar, T)}{\alpha_r({\rm X}^{i}, T)} ~~~;
\ee 
if $x \equiv n_e/\nH \le x_{\rm crit}$, 
then grain-assisted removal
of X$^i$ occurs more rapidly than radiative 
recombination to X$^{i-1}$.  We use the FORTRAN routine rrfit.f
to evaluate radiative recombination 
coefficients.\footnote{\label{fn:verner} The subroutine  
	rrfit.f was written by D.~A.~Verner and is available on the World
	Wide Web at \url{www.pa.uky.edu/$\sim$verner/fortran.html}.}
In Figure \ref{fig:x_crit_cnm}, we take $T=100\K$, appropriate for the cold
neutral medium (CNM).  For a nominal CNM electron fraction 
$x \approx 10^{-3}$ and $\cpar \approx 400 \K^{1/2} \cm^3$, the grain-assisted 
ion removal rate is generally comparable to the radiative recombination 
rate and therefore will appreciably affect the ionization balance for
H$^+$ and most metal ions.

In Figure \ref{fig:x_crit_wnm}, we take $T=6000\K$, appropriate for the
warm neutral medium (WNM) and warm ionized medium (WIM).  Adopting 
$x \approx 0.1$ and 
$\cpar \approx 3000 \K^{1/2} \cm^3$ for the WNM, we find that
the grain-assisted ion removal rate $\ltsim 10\%$ the radiative 
recombination rate.  
For the WIM ($x \approx 1$, $\cpar \approx 1000 \K^{1/2} \cm^3$), 
grain-assisted ion removal contributes at most at the few percent level.

\section{\label{sec:n_e} Electron density determinations including 
grain-assisted ion removal}

When grain-assisted ion removal is important, the ionization balance 
equation (\ref{eq:ionization_equil}) must be replaced with the following 
set of equations:
\be
\label{eq:2level_1}
\frac{dn_{i+1}}{dt} = \Gamma ({\rm X}^i) 
n_i - \alpha_r({\rm X}^{i+1}) n_e n_{i+1} - 
\alpha_g({\rm X}^{i+1}) \nH n_{i+1}
\ee
and
\be
\label{eq:2level_2}
\frac{dn_i}{dt} = - \Gamma({\rm X}^i) 
n_i + \alpha_r({\rm X}^{i+1}) n_e n_{i+1} + 
\alpha_g({\rm X}^{i+1}) \nH n_{i+1} (1-s)~~~,
\ee
where $s$ is the probability that an ion sticks to a grain following charge 
exchange.\footnote{We assume that ions do not stick to grains following 
collisions without charge exchange.}  Of course, if more than two 
ionization stages are populated, then more equations are needed.
When 
$s>0$, $n_i$ and $n_{i+1}$ both decay, but, with the exception of transients
that die out rapidly, the ratio $n_{i+1}/n_i$ is constant:
\be
\label{eq:n_1/n_0}
\frac{n_{i+1}}{n_i} \approx \frac{q}{2b} \left[ 1+ \left( 1+\frac{4b}{q^2}
\right)^{1/2} \right]~~~,
\ee
where $b \equiv r_1+r_2 (1-s)$, $q \equiv 1-r_1-r_2$, 
$r_1 \equiv \alpha_r n_e/\Gamma$, and 
$r_2 \equiv \alpha_g \nH / \Gamma$.  

Solving equation (\ref{eq:n_1/n_0}) for $n_e$ in terms of the observed 
column density ratio $R \equiv n({\rm X}^i)/n({\rm X}^{i+1})$, we find
\be
\label{eq:n_e}
n_e = \frac{R \Gamma}{\alpha_r} \left[ 1 - \frac{1-s+R}{1+R} \frac{\alpha_g
\nH}{R \Gamma} \right]~~~.
\ee
Since $\alpha_g$ depends on $n_e$ (through its dependence on the charging
parameter $\cpar$), equation (\ref{eq:n_e}) must be solved
iteratively, but generally converges rapidly.
Usually $r_1\ll 1$ and $r_2 \ll 1$ (implying $R \ll 1$).  Thus, the 
impact of
grain-assisted ion removal on the electron density determination decreases
as the sticking probability $s\rightarrow1$.  The observed 
interstellar depletions vary widely from element to element, implying 
$s \ll 1$ for some elements (e.g.~S) while $s \sim 1$ for others 
(e.g.~Ti, Ca; see Weingartner \& Draine 1999).  This wide variation in $s$
could also lead to discrepancies in electron density estimates when 
different elements are observed.

A large fraction of the Ca in diffuse clouds can be in Ca$^{++}$; thus, 
we must consider the ionization equilibrium of a 3-level system in this 
case.  For arbitrary $s$, this results in a rather complicated cubic 
equation for $n_e$ in terms of the observed ratio $R_1 \equiv n({\rm Ca}^0)/
n({\rm Ca}^+)$.  However, the equation simplifies to a quadratic when $s=0$ 
or $s=1$.  Before displaying the results, it is convenient to 
introduce some notation:  $\tilde{n}_e \equiv R_1 \Gamma({\rm Ca}^0)/
\alpha_r({\rm Ca}^+)$ (this is the value of $n_e$ when grains are 
unimportant), 
$g_1 \equiv \alpha_g({\rm Ca}^+)/\alpha_r({\rm Ca}^+)$, 
$g_2 \equiv \alpha_g({\rm Ca}^{++})/\alpha_r({\rm Ca}^{++})$,
$d \equiv \alpha_r({\rm Ca}^{++})/\alpha_r({\rm Ca}^+)$, 
$\gamma \equiv \Gamma({\rm Ca}^+)/\Gamma({\rm Ca}^0)$,
$h_1 \equiv \alpha_g({\rm Ca}^+) \nH / \alpha_r({\rm Ca}^+) \tilde{n}_e$,
$h_2 \equiv \alpha_g({\rm Ca}^{++}) \nH / \alpha_r({\rm Ca}^+) \tilde{n}_e$,
and $f_2 \equiv f_{{\rm Ca}^{++} \rightarrow {\rm Ca}^0}$
(from eq.~\ref{eq:f_2}).

When $s=0$,
\be
\label{eq:3-level_s=0}
n_e = \frac{1}{2}(\tilde{n}_e -g_1 \nH) (1+Q_0) -\frac{1}{2} g_2 \nH
(1-Q_0)~~~,
\ee
where
\be
Q_0 = \left\{ 1 - \frac{4 g_2 \nH \gamma f_2 \tilde{n}_e}{R_1 [(g_1 - g_2)
\nH - \tilde{n}_e]^2}\right\}^{1/2}~~~.
\ee
Note that equation (\ref{eq:3-level_s=0}) reduces to equation 
(\ref{eq:n_e}) when $\gamma =0$ or $f_2=0$.
The ratio $R_2 \equiv n({\rm Ca}^0)/n({\rm Ca}^{++})$ is given by
\be
R_2 = \frac{R_1^2}{\gamma} \left( h_2 + d \frac{n_e}{\tilde{n}_e} \right)~~~.
\ee

When $s=1$,
\be
\label{eq:3-level_s=1}
n_e = \tilde{n}_e \left[ 1 - \frac{R_1 h_1 (1+d R_1) (1+Q_1) + R_1 (d+h_2)
(1+R_1) (1-Q_1) + \gamma}{2 (1+R_1) (1+d R_1)} \right]~~~,
\ee
where
\be
Q_1 = \left( 1 + \frac{\gamma^2 + 2 \gamma R_1 \left\{ h_1 (1+d R_1) + 
(1+R_1) [d-h_2 (1+2 d R_1)]\right\}}{R_1^2 [ h_1 (1+d R_1) - (d+h_2)
(1+R_1)]^2} \right)^{1/2}~~~;
\ee
\be
\label{eq:n_e_3_last}
R_2 = \frac{R_1^2}{\gamma} \left[ h_2 - R_1^{-1} + \left( d + R_1^{-1}
\right) \frac{n_e}{\tilde{n}_e} \right]~~~.
\ee
Note that equation (\ref{eq:3-level_s=1}) reduces to equation (\ref{eq:n_e})
when $\gamma =0$.

\section{\label{sec:observations} Application to Observations}

\subsection{\label{sec:FS} HD 215733}

\begin{figure*}[tbh]
\centerline{\epsfig{file=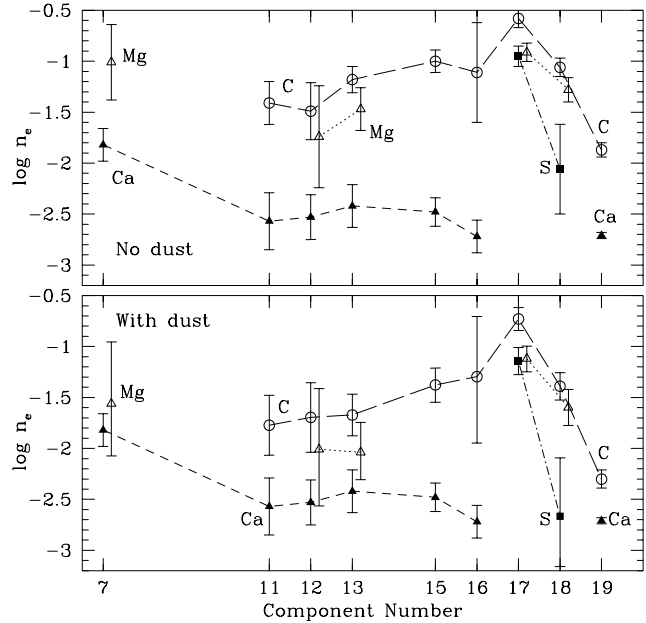,width=\fullfigurewidth}}
\caption{
\label{fig:FS97}
Upper panel:  Fitzpatrick \& Spitzer (1997) results for $n_e$ (in $\cm^{-3}$)
for cool 
cloud components along the line of sight to HD~215733, with $n_e$
determined from 
ionization equilibrium for Ca$^+$, C$^0$, Mg$^0$, and S$^0$ (Mg results are
adjusted as per Fitzpatrick 1997).
Lower panel:  
Modified $n_e$ determinations, with grain-assisted ion removal included.
We assume $s=1$ for Ca and $s=0$ for C, Mg, and S.  
        }
\end{figure*}

Fitzpatrick \& Spitzer (1997) performed ionization equilibrium analyses of
Ca$^+$/Ca$^{++}$, C$^0$/C$^+$, Mg$^0$/Mg$^+$, and S$^0$/S$^+$
for several clouds along the line of sight to HD~215733.  
In the case of Ca$^+$, the Ca$^{++}$ abundance was estimated based on
an assumed gas-phase Ca abundance.
In Table \ref{tab:FS97}, we reproduce their inferred temperatures
and H number densities for the cold cloud components.  In the upper 
panel of Figure \ref{fig:FS97}, we reproduce their electron density 
determinations.\footnote{The Mg$^+$ f-values used by Fitzpatrick \& 
Spitzer (1997) are apparently too large by a factor of $\approx 2.4$ 
(Fitzpatrick 1997).  Thus, we have multiplied the $n_e$ values obtained 
from the Mg column density ratio by 2.4.}
The electron densities inferred from the C and Mg abundance ratios 
appear to be roughly consistent, but those inferred from the Ca abundance 
ratios are systematically lower.  

In the lower panel of Figure \ref{fig:FS97}, we display revised
$n_e$ determinations that take grain-assisted ion removal into account. 
We assume $s \approx 1$ for Ca, which is observed to be highly depleted in
cold clouds, and $s \approx 0$ for C, Mg, and S, which are not as severely 
depleted.  With this modification, the Mg determinations lie between the 
C and Ca determinations, and are consistent with both.  The C and Ca 
determinations remain inconsistent.  If dissociative recombination of 
CH$^+$ is important in these clouds, then the $n_e$ inferred from the 
C abundance ratios are too high; unfortunately, the CH$^+$ column density
towards HD 215733 is not known.  Alternatively, the $n_e$ inferred from the
Ca abundance ratios could be too small,
as would be the case if the Ca abundance has been overestimated.
Unfortunately, with only three elements for which both ionization stages
have been observed, it is difficult to arrive at definitive conclusions
regarding the electron densities in many of these components.

\subsection{\label{sec:23Ori} 23 Ori}

\begin{figure*}[tbh]
\centerline{\epsfig{file=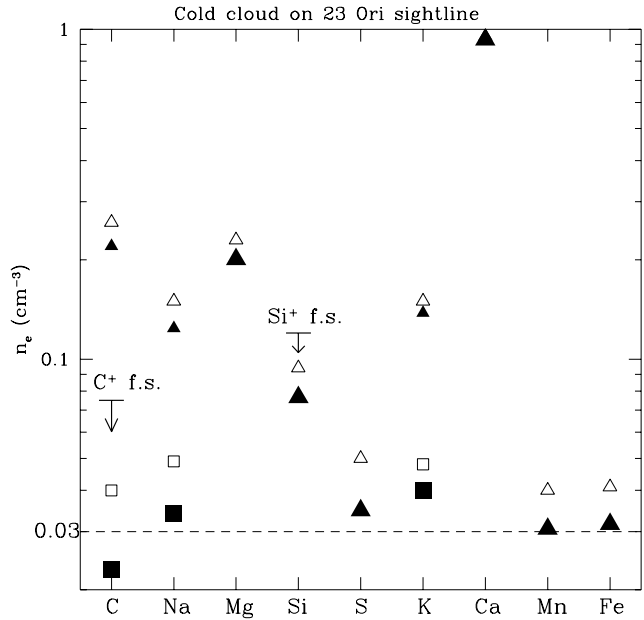,width=\fullfigurewidth}}
\caption{
\label{fig:welty}
Electron density $n_e$ inferred from ionization equilibrium for the cold gas
towards 23 Orionis.  Open triangles:  Welty et al.~(1999) analysis; 
Filled triangles:  Welty et al.~(1999) analysis plus grain-assisted ion
removal;  Open squares:  Modified analysis (see text), without 
grain-assisted ion removal;  Filled squares:  Modified analysis, 
including grain-assisted ion removal.  For each species, our favored $n_e$
determination is indicated by an enlarged symbol.  Our actual favored
$n_e \approx 0.03 \cm^{-3}$ is indicated by the dashed line.  Upper limits
from the observed fine structure excitation from C$^+$ and Si$^+$ are 
also indicated.
}
\end{figure*}

Welty et al.~(1999, hereafter W99) have conducted a detailed investigation 
of the diffuse
clouds along the line of sight to 23 Orionis.  They inferred $T\sim 100\K$,
$\nH \sim 10 \cm^{-3}$, and $N_{\rm H} \approx 5 \times 10^{20} \cm^{-2}$ 
for the cold cloud material along this sightline
at heliocentric velocities $20 < v_\odot < 27\kms$.  They
also inferred $n_e$ using equation (\ref{eq:ionization_equil}) for 
ionization equilibrium (between X$^0$ and X$^+$)
with several different elements, and found widely varying values; we display
their electron densities as open triangles in Figure 
\ref{fig:welty}.\footnote{We display the W99 $n_e$ determinations using
photoionization rates for the ``WJ1'' radiation field (see Table
\ref{tab:ip_r_0}); these photoionization rates are very similar to those
for the ISRF of Mathis et al.~(1983).}  
The electron densities inferred if ionization equilibrium is assumed
vary over about an order of magnitude, depending on which element is used.

Electrons will collisionally excite the $^2$P$_{3/2}$
excited fine structure levels of
C$^+$ and Si$^+$.
W99 report $N({\rm C}^+)=10^{16.95}\cm^{-2}$, 
and $N({\rm C}^{+*})<10^{15.0}\cm^{-2}$, from which 
we infer $n_e < 0.075\cm^{-3}$; we assumed $T=100\K$,
a collision strength $\Omega=2.1$ from Keenan et al. (1986)
and H collisional rates from Launay \& Roueff (1977).

For Si$^+$, W99 find $N({\rm Si}^+)=10^{15.37}\cm^{-2}$ and
$N({\rm Si}^{+*})<10^{10.7}\cm^{-2}$, from which we find
$n_e < 0.12\cm^{-3}$,
assuming a collision strength $\Omega=5.74$ (Keenan et al. 1985). 
H collisional rates have not
been calculated; we use the 
rate coefficient calculated by Launay \&
Roueff for deexcitation of C$^+$($^2$P$_{3/2}$).

Upper limits on $n_e$ derived from fine-structure excitation
are shown in Figure \ref{fig:welty}.

The electron densities inferred from ionization equilibrium 
including grain-assisted ion removal are 
displayed as filled triangles in
Figure \ref{fig:welty}.  We have assumed $s=0$ in order to maximize
the effect that grain-assisted ion removal might have.  
(For most of the elements in Figure \ref{fig:welty}, this assumption seems
reasonable.  For Fe and Ca, however, the large depletions observed in 
interstellar gas suggest that $s \gtsim 0.1$.)
The W99 electron densities inferred using C, Na, Mg, and K are particularly 
large; the resulting $n_e/\nH$ are high enough that the inclusion of 
grain-assisted ion removal
does not dramatically alter the inferred values of $n_e$.
Thus, the various $n_e$ determinations do not appear to be reconciled by
including this process.

However, some of the W99 $n_e$ determinations might be too high
for other reasons.  Dissociative recombination of 
CH$^+$ might be an important additional recombination mechanism for C. 
W99 found that if the CH$^+$ observed towards 23 Ori resides in 
the cold cloud for which the C$^0$ and C$^+$ 
column densities were measured, then the estimated $n_e$ must be revised 
downward from 0.26 to $0.04 \cm^{-3}$ (open square in Figure 
\ref{fig:welty}); of course, grain-assisted ion removal would reduce 
it further (filled square). 

Na$^+$ and K$^+$ are unobservable since they have no resonance lines below
the Lyman limit.  Thus, W99 had to assume gas-phase Na and K abundances  
in order to analyze the ionization balance for these elements.
W99 assumed depletions of 0.5 dex (relative to solar) for both elements.
If, instead, these elements are assumed to be undepleted, then the $n_e$
estimates are substantially smaller; these are indicated in Figure
\ref{fig:welty} by filled (open) squares for analyses that do (do not)
include grain-assisted ion removal.  

Note that Welty \& Hobbs (2001) argue, on the basis of a large number of 
observed sight lines, that Na and K are generally depleted from the 
gas phase by 0.6--0.7 dex relative to solar.  
We consider the depletion estimates for Na and K to be quite uncertain,
as they rely on knowledge of the electron density (which, as we have
seen, is uncertain since ``standard'' ionization analyses give discrepant
results when multiple elements are used).

If the revisions described above (for C, Na, and K) are correct, then the 
observed column densities for six elements (C, Na, S, K, Mn, Fe)
yield $n_e \approx 0.03 \cm^{-3}$, consistent with the 
upper limits $n_e < 0.075 \cm^{-3}$ and $n_e < 0.12 \cm^{-3}$
obtained from the fine structure excitation of C$^+$ and Si$^+$.
For $n_e/\nH=3\times10^{-3}$, grain-assisted ion removal
contributes significantly to the recombination.  

However, discrepancies remain for three elements: Ca, Mg, and Si.
\begin{enumerate}
\item
	W99 found $n_e \approx 0.95 \cm^{-3}$ from 
	Ca$^0$/Ca$^+$ ionization balance.
	Using equation (\ref{eq:3-level_s=0}),
	we find $n_e = 0.93 \cm^{-3}$ even 
	if $s=0$, so including recombination on grains
	does not resolve this discrepancy.
\item
	Even after allowing for grain recombination with $s=0$,
	the Mg$^0$/Mg$^+$ abundance ratio implies an electron density 
	$n_e\approx0.2\cm^{-3}$---a
        factor of 6 greater than our best estimate.
\item
	Even after allowing for grain recombination with $s=0$,
	the Si$^0$/Si$^+$ abundance ratio implies an electron density
	$n_e\approx 0.08\cm^{-3}$---a factor of 2.5 greater than our
	best estimate.
\end{enumerate}

Unless the photoionization rates have for some reason been overestimated, 
the observations suggest that additional physical processes are involved
for these three elements.  

\section{\label{sec:additional_processes} Possible Additional Processes 
Affecting Ionization Balance in 23 Ori}

\subsection{Dust Destruction?}

\begin{figure*}[tbh]
\centerline{\epsfig{file=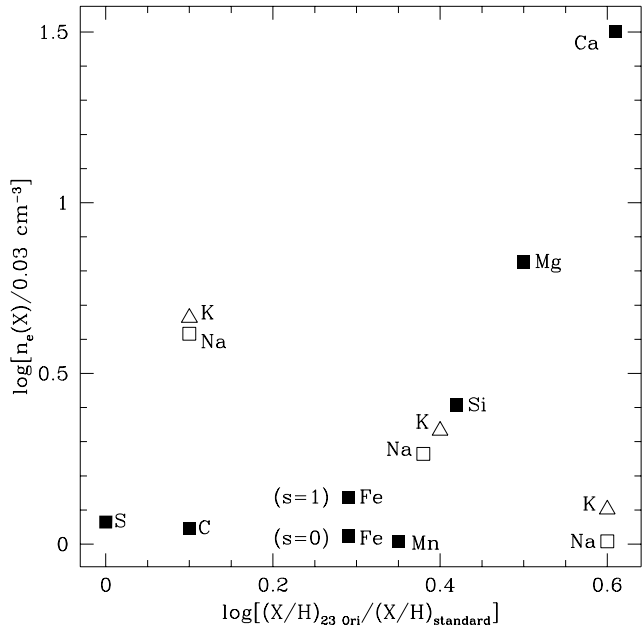,width=\fullfigurewidth}}
\caption{
\label{fig:deplete}
Electron density for the cold cloud along the line of sight to 23 Ori
inferred from observed value of $N(X^0)/N(X^+)$, 
normalized to our favored value of $0.03 \cm^{-3}$, versus the factor
by which the gas-phase element abundance is observed to be
enhanced above its ``standard'' cold
cloud value.  The Na and K depletions are unknown, so for these elements 
we display a few possible points.  The location of the Mn point is  
uncertain since W99 report only a marginal detection of Mn$^0$.
        }
\end{figure*}

In Figure \ref{fig:deplete}, we plot
the inferred electron density, normalized to our favored value of 
$0.03 \cm^{-3}$, versus the gas-phase element abundance in the cold cloud 
towards 23 Ori, normalized to the standard cold cloud abundance (see Table 5
in W99).  We have included two points for Fe, with sticking coefficient 
$s=0$ and $s=1$; Fe is generally strongly depleted in cold clouds, 
suggesting that the sticking coefficient may be $\sim 1$.  
Figure \ref{fig:deplete} shows a clear trend: greater 
neutral abundance enhancements are observed for elements 
with greater overall abundance enhancements.  The total gas-phase abundances
of Na and K are not known, but if they are enhanced by $\approx 0.4 \,$dex over
their standard values, then they fit the trend nicely.  (We have also 
included points for Na and K for which the gas-phase abundance is enhanced
by 0.1 dex, as assumed by W99, and by 0.6 dex, as assumed by us above.
Neither of these choices lie close to the trend.)  The observed correlation 
suggests the possibility that the enhanced neutral Mg, Si, and Ca abundances
result from ongoing destruction of dust containing these elements, 
with the atoms
injected into the gas as neutrals.  However, the production rate of neutrals
due to dust destruction would have to exceed the rate due to recombination;
it is hard to imagine how such rapid dust destruction could be occurring.
It seems to us unlikely that dust destruction is affecting the ionization
balance observed toward 23 Ori.

\subsection{Dielectronic Recombination?}

It is intriguing to note that 
Mg$^0$ ($3s^2$ ground state), Si$^0$ ($3p^2$ ground state),
and Ca$^0$ ($4p^2$ ground state)
are all species for which dielectronic recombination is important for
$T < 10^4\K$; the dielectronic recombination rate for 
${\rm Mg}^+\rightarrow{\rm Mg}^0$
exceeds the radiative recombination rate for $T> 5800\K$.
W99 argue for a temperature $T\approx 100\K$ for this velocity
component, but perhaps
there is some gas in this velocity range with $5000 \ltsim T \ltsim 10^4\K$
and a relatively high electron density, leading to preferential recombination
of those species with the largest (radiative + dielectronic) 
rate coefficients at these temperatures.
It is also intriguing to note that W99 report detection of Al$^{++}$
at the velocity of the dominant neutral absorption, suggesting that
some of the material in this velocity range may be photoionized by photons 
beyond the Lyman limit, with accompanying ionization of H and presumably
enhanced values of $n_e$ and temperatures $T\approx10^4\K$.
Recall that CH$^+$ is present in this velocity range.
The origin of interstellar CH$^+$ remains unclear, but all current scenarios
for CH$^+$ formation, whether in
MHD shock waves (Draine \& Katz 1986; Pineau des For\^ets et al.\ 1986),
turbulent boundary layers of clouds (Duley et al 1992),
strong Alfv\'en waves (Federman et al.\ 1996),
or intermittent dissipation of turbulence
(Falgarone, Pineau des For\^ets, \& Roueff 1995; Joulain et al.\ 1998),
involve regions of enhanced temperature.

We have found that if $\approx 10\%$ of the 
column density arises in a warm component with $T \approx 10^4 \K$ and 
$n_e \approx 4 \cm^{-3}$, then the predicted column density ratios for all
of the elements in Figure \ref{fig:welty} are within a factor of $\approx 2$
of the observed ratios.  In this case, however, the predicted 
populations of the excited
fine structure levels of C$^+$ and Si$^+$ substantially exceeds the 
observational upper limits
(see \S 3.3.3 of W99).  We were unable to find any models for
which (a) dielectronic recombination in warm gas significantly increases the 
neutral/ionized column density ratios of Mg, Si, and Ca and (b)  
fine structure levels are not overpopulated.
We conclude that dielectronic recombination cannot explain the observed high
values of Mg$^0$/Mg$^+$, Si$^0$/Si$^+$, and Ca$^0$/Ca$^+$.\footnote{W99
also conclude that dielectronic recombination is not likely to be 
responsible for the enhanced Ca$^0$/Ca$^+$.  They note that the $n_e$ 
derived from Ca$^0$/Ca$^+$ are often higher than those derived from other
neutral/ionized ratios, but that the line widths of Ca$^0$ usually indicate
an origin in cool gas.}

\subsection{Chemistry?}


W99 suggested that $n_e \approx 0.15 \cm^{-3}$ in the cold gas towards 23 Ori 
and that charge exchange with protons might be an important ionization 
mechanism for S, Mn, and Fe, accounting for the lower $n_e$ inferred from 
these species.  
W99 noted that charge transfer rates for $T\approx100\K$ have not
been calculated for these elements with protons, although potential
energy curves calculated for SH$^+$ (Kimura et al. 1997) imply that the
rate coefficient for ${\rm S} + {\rm H}^+ \rightarrow {\rm S}^+ + {\rm H}$
is very small at low temperatures, so that protons would not affect the
S$^0$/S$^+$ ionization balance.
It should also be noted that with the limit $n_e \ltsim 0.075 \cm^{-3}$
from C$^+$ fine structure excitation, the charge transfer ionization
rates $n({\rm H}^+)\langle\sigma v\rangle \ltsim 2\times10^{-10}\s^{-1}$
for Mn and Fe
since $\langle\sigma v\rangle\ltsim 3\times10^{-9}\cm^3\s^{-1}$ even if 
MnH$^+$ or FeH$^+$ have favorable potential energy curves.
Charge transfer with protons is therefore at most comparable to photoionization
for Mn and Fe (see Table \ref{tab:ip_r_0}), so that $n_e$
inferred from Mn and Fe cannot be raised by more than a factor
of 2 even if charge transfer with protons is maximally effective.
We therefore believe that the electron densities inferred from S, Mn, and
Fe are reliable.

It is intriguing to note that the three atoms with $np^2$ ground states
[C($2p^2$), Si($3p^2$), and Ca($4p^2$)] are all overabundant.
In the case of C we can attribute the overabundance to dissociative
recombination of CH$^+$, and the question arises whether
dissociative recombination of SiH$^+$ and
CaH$^+$ might also be taking place.
CH$^+$ is believed to be produced by the exchange reaction
${\rm C}^+ + {\rm H}_2 + 0.40 {\rm eV} \rightarrow {\rm CH}^+ + {\rm H}$.
The reaction $X^+ + {\rm H}_2 \rightarrow X{\rm H}^+ + {\rm H}$
is much more endothermic for Si$^+$ (1.29 eV) than for C$^+$ (0.40 eV).
While the scenario for CH$^+$ formation remains unclear, the increased
endothermicity for Si (and presumably for Ca, though the heat of formation of
CaH$^+$ is unavailable) makes such a chemical explanation appear unlikely.
Furthermore, production of SH$^+$ would appear more favorable than, say,
SiH$^+$, yet the observed S$^0$/S$^+$ ratio can be understood without
invoking dissociative recombination of SH$^+$.

\subsection{Atomic Physics?}

The overabundance of the $np^2$ species suggests that some
common atomic physics might be involved.  If, for example, photoionization
rates have been overestimated, or radiative recombination rates at
$T\approx 100\K$ have been underestimated, the rates for all three elements
(C, Si, Ca) might be similarly affected.  
The photoionization rates are probably unlikely to be seriously in error,
but possible 
underestimation of the low temperature radiative recombination rates
should be considered.
The Milne relation for the rate coefficient for radiative recombination
$X^+ + e \rightarrow X + h\nu$ is
\begin{equation}
\label{eq:milne}
\alpha = \frac{4\pi}{(2\pi m_e kT)^{3/2}}
\frac{1}{g_{X^{\!+}} c^2}
\int_0^\infty dE e^{-E/kT}
\sum_j g_j (E+I_j)^2\sigma_{pi,j}(h\nu=I_j+E)~~~,
\end{equation}
where the sum over $j$ is over all of the terms for the ground configuration
(e.g., $2s^22p^2\,^3$P, $^1$D, $^1$S for C)
and 
all of the terms for one-electron excited configurations of the atom
(e.g., $2s2p^3\,^5$S, $2s^22p3s\,^3$P, $2s^22p3s\,^1$P, $2s2p^3\,^3$D, 
...  for C),
$g_{X^{\!+}}$ is the degeneracy of the ion $X^{\!+}$,
$g_j$ is the degeneracy of excited state $j$ of the atom $X$,
$I_j$ is the energy required to ionize from excited state $j$,
and $\sigma_{pi,j}(h\nu)$ is the photoionization cross section from
excited state $j$.
Equation (\ref{eq:milne}) shows that the radiative recombination rate
at $T\ltsim 100\K$ could be large if some state $j$ has a photoionization
cross section $\sigma_{pi,j}$ with a resonance 
within $\ltsim 0.01\eV$ of threshold.
While we have no indication that such a resonance exists, we encourage
atomic physicists to reexamine the calculated photoionization rates and
implied radiative recombination rates.
Such a near-threshold photoionization resonance, and consequent large
low-temperature recombination rate,
could conceivably also occur for Mg$^0$.

\section{\label{sec:x_e} Electron fraction in the cold neutral medium}

In this section, we investigate the extent to which grain-assisted removal
of H$^+$ and He$^+$ decreases the electron fraction $x_e$
in the CNM.  There are two separate contributions to the
total electron fraction.  The first is due to the ionization of metals, 
primarily C, by the far-ultraviolet radiation field.  Since we expect C
to always be primarily in C$^+$ in cold diffuse clouds, we simply set 
this contribution to $x_{e,0} \approx 2 \times 10^{-4}$.  The second
contribution is due to ionization (primarily of H and He) by cosmic rays
and by the diffuse X-ray radiation field.  

Wolfire et al.~(1995) estimate the cosmic ray ionization rate for H and He 
(including ionization due to secondary electrons) in diffuse clouds to be 
$\nH \xi_{\rm CR} \approx 3 \times 10^{-17} \nH \s^{-1}$ (see their 
\S 2.2.2).  They also give an approximate expression for the H and He X-ray 
ionization rate $\xi_{\rm XR}$ (for the X-ray radiation field in the local 
neighborhood) 
in terms of $x_e$ and the column density $N_w$ of warm gas 
surrounding a cold diffuse cloud (see their Appendix A).  Following 
Wolfire et al., we adopt $N_w=10^{19} \cm^{-2}$.  

For simplicity, we assume that all of the ionization resulting from cosmic
rays and X-rays comes from H;\footnote{ 
	The rates for radiative recombination and grain recombination
	for He$^+\rightarrow \,$He are not very
	different from those for H$^+\rightarrow \,$H, so the derived
	electron density is not very sensitive to the fraction of the 
	ionizations that ionize H vs.~He.  A treatment including He would
	likely result in $\ltsim 10\%$ increases in $n_e$.} 
in this case, $x_e = x_{e,0} + \xH$, where
$\xH \equiv n({\rm H}^+)/\nH$.  Ionization equilibrium for H is expressed 
by the following equation:
\be
\label{eq:ionization_H}
(\xi_{\rm CR} + \xi_{\rm XR}) \nH (1-\xH) = (\alpha_r({\rm B}) x_e + \alpha_g)
\xH n_{\rm H}^2~~~.
\ee
Here we use the ``case B'' recombination coefficient; i.e., only recombinations
to states with principal quantum number $n \ge 2$ are included, since 
recombinations to $n=1$ result in an ionizing photon which is immediately
absorbed by a nearby H atom.  We take $\alpha_r({\rm B}) = 3.5 \times
10^{-12} (T/300 \K)^{-0.75} \cm^3 \s^{-1}$ (Liszt 2001).

\begin{figure*}[tbh]
\centerline{\epsfig{file=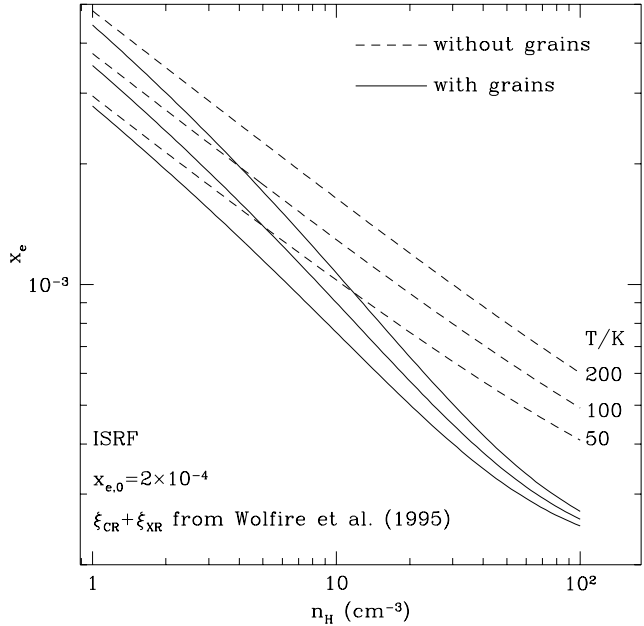,width=\fullfigurewidth}}
\caption{
\label{fig:x_ISM}
Electron fraction $x_e$ as a function of the H density $\nH$.  The 
electron fraction consists of a term $x_{e,0}\approx 2 \times 10^{-4}$ due
to the ionization of metals (primarily C) by the far-ultraviolet radiation 
field and a term due to the ionization of H and He by cosmic rays and X-rays.
Higher curves are for higher gas temperatures $T$, as indicated.  The solid
(dashed) curves were computed with (without) the grain-assisted removal of
H$^+$ and He$^+$.
        }
\end{figure*}

In Figure \ref{fig:x_ISM}, we display the resulting electron fraction as a
function of $\nH$ for a range of temperatures characteristic of cold diffuse
clouds (solid curves).   
We also show the electron fraction that would result if 
grain-assisted ion removal were not included in the ionization balance
(dashed curves).  As $\nH$ increases, grain-assisted ion removal becomes
more important and the results for $x_e$ diverge.  When $\nH=30\cm^{-3}$
and $T=100\K$, grain-assisted ion removal reduces $x_e$ from 
$\approx 8.0 \times 10^{-4}$ to $\approx 4.5 \times 10^{-4}$, a 45\%
reduction.  

\section{\label{sec:conclusions} Conclusions}

The principal results of this paper are as follows:
\begin{enumerate}

\item We define a rate coefficient $\alpha_g$
for the removal of ions from the gas phase
due to charge exchange with grains, including PAHs (eq.~\ref{eq:alpha_g});
``removed'' ions either remain in the gas phase, in a lower ionization stage,
or stick to the grain.  We compute $\alpha_g$, for several astrophysically 
important ions, as a function of the gas temperature $T$ and grain-charging
parameter $\cpar$ and provide a convenient fitting formula 
(eq.~\ref{eq:alpha_g_approx} and Table \ref{tab:alpha_g}).  

\item We evaluate a critical electron fraction $x_{\rm crit}$;
if $x \equiv n_e/\nH < x_{\rm crit}$, then grain-assisted ion removal occurs
more rapidly than radiative recombination.  We find that the rates for
grain-assisted ion removal and radiative recombintion are comparable in 
the CNM, but that radiative recombination dominates in the WNM and WIM
(Figs.~\ref{fig:x_crit_cnm} and \ref{fig:x_crit_wnm}).

\item We provide formulae for deriving the electron density $n_e$ from 
observed neutral/ionized column density ratios when grain-assisted ion 
removal is important, for the cases that only two ionization stages are
significantly populated (eq.~\ref{eq:n_e}) and that three stages are 
populated (eqs.~\ref{eq:3-level_s=0}--\ref{eq:n_e_3_last}).

\item We have investigated whether or not grain-assisted ion removal 
can reconcile various estimates obtained by Spitzer \& Fitzpatrick (1997)
for $n_e$ in the cold gas clouds along the line of sight to HD 215733
(\S \ref{sec:FS}).  Without grain-assisted ion removal, the C and Mg 
determinations are consistent, while the Ca determinations are systematically
lower.  When grain-assisted ion removal is included, the Mg determinations
are intermediate between the C and Ca determinations, and consistent with 
both.  However, the C and Ca determinations remain inconsistent.

\item We have investigated whether or not grain-assisted ion removal can 
reconcile the various estimates obtained by W99 for $n_e$ in the cold gas
along the line of sight to 23 Ori (\S \ref{sec:23Ori}) 
W99 estimated $n_e$ by observing 
neutral/ionized column density ratios for several elements and assuming that
only ionization by the interstellar radiation field and radiative 
recombination are important to the ionization equilibrium.  If 
grain-assisted ion recombination is the only additional process relevant
to the ionization equilibrium, then the various $n_e$ determinations are 
not reconciled.  However, dissociative recombination of CH$^+$ may be
an important recombination mechanism for C, and the Na and K
neutral/ionized ratios
inferred by W99 may be incorrect, if their assumed depletions are incorrect.
Making use of these degrees of freedom, we find that most of the observed
abundance ratios are consistent with $n_e \approx 0.03 \cm^{-3}$.  However,
Mg, Si, and Ca remain discrepant, with inexplicably high neutral/ionized 
ratios.  We have considered the possibility that
the observed neutral/ion ratios for Mg, Si, and Ca might be due to an
additional warm gas component with $T\approx10^4\K$ and $n_e\gtsim 3\cm^{-3}$.
However, such a scenario does not appear to be consistent with the observed
C$^+$ and Si$^+$ fine structure excitation.  
Alternatively,
the enhanced abundances of Mg$^0$, Si$^0$, and Ca$^0$ could
perhaps be explained if dust is being rapidly destroyed and injecting 
neutral Mg, Si, and Ca into the gas, but it is hard to imagine how the 
dust destruction could occur rapidly enough for this to be the case.  

\item We suggest that low temperature radiative recombination rates
might have been underestimated for Mg, Si, and Ca.

\item We have found that the grain-assisted removal of H$^+$ and He$^+$
can have a significant effect on the electron fraction in the CNM
(\S \ref{sec:x_e}).  For clouds with $T\approx 100 \K$ and 
$\nH \approx 30 \cm^{-3}$, $x$ is reduced from $8.0 \times 10^{-4}$ to 
$4.5 \times 10^{-4}$.  We have assumed that the ionization of C contributes
an electron fraction of $2 \times 10^{-4}$.

\end{enumerate}

\acknowledgements
This research was supported in part by NSF grant
AST-9988126 and by an NSF International Research Fellowship to J. C. W.
We thank H. Liszt, T. M. Tripp, and P. A. M. van Hoof 
for helpful discussions.
We especially thank E. B. Jenkins and D. E. Welty
for a number of valuable suggestions.
We are grateful to R. H. Lupton for the availability of the SM plotting 
package.

\begin{appendix}
\section{Grain-Ion Interaction Potential}

We assume that the electrostatic potential energy of an ion of charge
$ze$ at the
surface of a grain with charge $Ze$ can be approximated by a point charge
$ze$ a distance $r_0$ from the surface of a conducting sphere with radius
$a$ and charge $Ze$ (see eq.\ [2.9] of Jackson 1962):
\begin{equation}
U(Z,z)=\frac{Zze^2}{a+r_0} + \frac{z^2e^2a^3}{2r_0(a+r_0)^2(2a+r_0)}
\end{equation}
Then,
the change in electrostatic interaction energy due to the electron
transfer from grain to ion is just $\Delta U(Z,z) = U(Z+1,z-1)-U(Z,z)$
(see eq.\ [\ref{eq:DeltaU}]).
\end{appendix}

\begin{deluxetable}{ccccccc}
\tablewidth{0pc}
\tablecaption{Atomic Data
\label{tab:ip_r_0}}
\tablehead{
\colhead{Species}&
\colhead{IP\tablenotemark{a}}&
\colhead{$r_0$\tablenotemark{b}}&
\colhead{$\Gamma_1$\tablenotemark{c}}&
\colhead{$\Gamma_2$\tablenotemark{d}}&
\colhead{$\alpha_{r+d}(T=100\K)$\tablenotemark{e}}&
\colhead{$\alpha_{r+d}(T=6000\K)$\tablenotemark{e}}
\\
\colhead{}&
\colhead{$\eV$}&
\colhead{$\Angstrom$}&
\colhead{$\s^{-1}$}&
\colhead{$\s^{-1}$}&
\colhead{$\cm^3 \s^{-1}$}&
\colhead{$\cm^3 \s^{-1}$}
}
\startdata
H$^0$  &13.60 &0.37 &...     &...       &8.61E-12  &6.01E-13\\
He$^0$ &24.60 &0.50 &...     &...       &8.42E-12  &6.71E-13\\
C$^0$  &11.26 &0.77 &2.0E-10 &3.3E-10   &8.63E-12  &7.44E-13\\
Na$^0$ &5.14  &1.86 &1.3E-11 &1.5E-11   &6.77E-12  &3.00E-13\\
Mg$^0$ &7.65  &1.60 &7.9E-11 &7.1E-11   &7.18E-12  &4.66E-13\\
Si$^0$ &8.15  &1.18 &3.0E-9  &3.1E-9    &9.39E-12  &1.56E-12\\
S$^0$  &10.36 &1.03 &7.8E-10 &1.2E-9    &7.46E-12  &5.66E-13\\
K$^0$  &4.34  &2.27 &5.6E-11 &6.3E-11   &1.11E-11  &4.16E-13\\
Ca$^0$ &6.11  &1.97 &3.7E-10 &4.3E-10   &7.07E-12  &2.40E-12\\
Mn$^0$ &7.43  &1.37 &1.4E-10 &1.5E-10   &8.33E-12  &2.16E-13\\
Fe$^0$ &7.90  &1.24 &2.0E-10 &2.0E-10   &8.52E-12  &2.48E-13\\
Ca$^+$ &11.87 &1.97 &1.4E-12 &2.5E-12   &2.70E-11  &1.02E-12\\
\enddata
\tablenotetext{a}{from Lide 2000}
\tablenotetext{b}{from Porile 1987}
\tablenotetext{c}{Ionization rate for the WJ1 radiation field, from 
P\'equignot \& Aldrovandi 1986}
\tablenotetext{d}{Ionization rate for the Draine radiation field, from
P\'equignot \& Aldrovandi 1986}
\tablenotetext{e}{Rate coefficient for radiative plus dielectronic 
recombination.  $\alpha_r$ from rrfit.f (see footnote \ref{fn:verner});
$\alpha_d$ from Aldrovandi \& P\'equignot 1974 (Na$^0$), Nussbaumer \&
Storey 1986 (Mg$^0$, Si$^0$), Shull \& van Steenberg 1982 (Ca$^0$, 
Ca$^+$), Arnaud \& Raymond 1992 (Fe$^0$).}
\end{deluxetable}

\begin{deluxetable}{cccccccc}
\tablewidth{0pc}
\tablecaption{Fitting Parameters for Grain-assisted Ion Removal Rate
Coefficient\tablenotemark{a}
\label{tab:alpha_g}}
\tablehead{
\colhead{Ion}&
\colhead{$C_0$}&
\colhead{$C_1$}&
\colhead{$C_2$}&
\colhead{$C_3$}&
\colhead{$C_4$}&
\colhead{$C_5$}&
\colhead{$C_6$}
}
\startdata
H$^+$     &3.436 &1.909E-6 &1.496 &1.790E3 &9.749E-4 &0.5612 &7.245E-5 \\
He$^+$ 	  &2.175 &2.031E-7 &2.114 &1.134E4 &2.197E-6 &1.0731 &9.119E-9 \\
C$^+$     &45.58 &6.089E-3 &1.128 &4.331E2 &4.845E-2 &0.8120 &1.333E-4 \\
Na$^+$    &2.178 &1.732E-7 &2.133 &1.029E4 &1.859E-6 &1.0341 &3.223E-5 \\
Mg$^+$    &2.510 &8.116E-8 &1.864 &6.170E4 &2.169E-6 &0.9605 &7.232E-5 \\ 
Si$^+$    &2.166 &5.678E-8 &1.874 &4.375E4 &1.635E-6 &0.8964 &7.538E-5 \\
S$^+$     &3.064 &7.769E-5 &1.319 &1.087E2 &3.475E-1 &0.4790 &4.689E-2 \\
K$^+$     &1.596 &1.907E-7 &2.123 &8.138E3 &1.530E-5 &1.0380 &4.550E-5 \\
Ca$^+$    &1.636 &8.208E-9 &2.289 &1.254E5 &1.349E-9 &1.1506 &7.204E-4 \\
Mn$^+$    &2.029 &1.433E-6 &1.673 &1.403E4 &1.865E-6 &0.9358 &4.339E-9 \\
Fe$^+$    &1.701 &9.554E-8 &1.851 &5.763E4 &4.116E-8 &0.9456 &2.198E-5 \\
Ca$^{++}$ &8.270 &2.051E-4 &1.252 &1.590E2 &6.072E-2 &0.5980 &4.497E-7 \\
\enddata
\tablenotetext{a}{See eq.~(\ref{eq:alpha_g_approx}).}
\end{deluxetable}

\begin{deluxetable}{ccc}
\tablewidth{0pc}
\tablecaption{Inferred Temperatures and Hydrogen Number Densities for Cold
Cloud Components Towards HD~215733\tablenotemark{a}
\label{tab:FS97}}
\tablehead{
\colhead{Component}&
\colhead{$T$}&
\colhead{$\nH$}
\\
\colhead{}&
\colhead{$\K$}&
\colhead{$\cm^{-3}$}
}
\startdata
7  &240 &22 \\
11 &100 &13 \\
12 &240 &5  \\
13 &110 &23 \\
15 &130 &21 \\
16 &100 &10 \\
17 &100 &18 \\
18 &50  &24 \\
19 &100 &10 \\  
\enddata
\tablenotetext{a}{From Fitzpatrick \& Spitzer 1997.}
\end{deluxetable}

\end{document}